\documentclass[referee,usegraphicx]{mn2e}

\def\beq{\begin{eqnarray}}
\def\eeq{\end{eqnarray}}

\def\a{\alpha}
\def\ac{\alpha_{con}}
\def\v{\upsilon}

\title[Radiatively inefficient accretion flows: advection versus convection]
{Global dynamics of radiatively inefficient accretion flows:
advection versus convection}

\author[Ju-Fu Lu, Shuang-Liang Li, and Wei-Min Gu]
{Ju-Fu Lu \thanks{E-mail: lujf@xmu.edu.cn},
Shuang-Liang Li and Wei-Min Gu \\
Department of Physics, Xiamen University, Xiamen 361005, China
}

\date{}

\begin{document}

\maketitle

\begin{abstract}
We obtain global solutions of radiatively inefficiently accretion flows
around black holes. Whether and where convection develops in a flow
are self-consistently determined with the mixing-length theory.
The solutions can be divided into three types according to the strength 
of normal viscosity. Type I solution corresponds to large viscosity 
parameter $\a \ga 0.1$, which is purely advection-dominated and with
no convection, and has been extensively studied in the literature.
Type II solution is for moderate $\a \sim 0.01$, which has a three-zone
structure. The inner zone is advection-dominated, the middle zone is 
convection-dominated and ranges from a few tens to a few thousands of
gravitational radii, and the outer zone is convectively stable and
matches outward a Keplerian disc. The net energy flux throughout the flow
is inward as in type I solution. Type III solution which is for small
$\a \la 0.001$ consists of two zones as Abramowicz et al. suggested
previously: an inner advection-dominated zone and an outer
convection-dominated zone, separated at a radius of a few tens of
gravitational radii. This type of solution has an outward net energy flux. 
In both type II and III solutions the radial density profile is between
the 1/2 law of self-similar convection-dominated accretion flow model and
the 3/2 law of self-similar advection-dominated accretion flow model, and
the efficiency of energy release is found to be extremely low. Our results
are in good agreement with those of recent numerical simulations.
\end{abstract}


\section{Introduction}

Accreting black holes in nearby galactic nuclei and low state X-ray
binaries are much dimmer than the standard Shakura-Sunyaev disc model
would predict. This phenomenon has been 
modeled within the framework of a radiatively inefficient 
accretion flow (RIAF). In such a 
flow, radiative losses are small because of the low particle 
density of the accreting plasma at 
low mass accretion rates. It was then suggested that most of 
the released gravitational and 
rotational energies of accreting plasma is advected inward in the 
form of the internal energy, 
and is finally absorbed by the black hole. Such a particular model 
of RIAFs was called the advection-dominated accretion flow (ADAF)
and attracted a considerable attention during the last decade
(see Narayan, Mahadevan \& Quataret 1998 and Narayan 2002 for reviews).

At the same time when the ADAF model was proposed, it was realized 
that ADAFs are 
likely to be convectively unstable in the radial direction because 
of the inward increase of the 
entropy of accreting gas (Narayan \& Yi 1994). Two dimensional (2D) 
and three dimensional 
(3D) hydrodynamical (HD) simulations of low viscosity RIAFs have confirmed 
the convective 
instability in these flows (Igumenshchev, Chen \& Abramowicz 1996;
Igumenshchev \& 
Abramowicz 1999, 2000; Stone, Pringle \& Begelmen 1999; 
Igumenshchev, Abramowicz \& 
Narayan 2000; McKinney \& Gammie 2002). Narayan, Igumenshchev \& 
Abramowicz (2000) 
and Quataert \& Gruzinov (2000) constructed another analytical 
model of RIAFs, which was 
based on a self-similar solution and reproduced the basic features 
of the HD 
simulations, and was called the convection-dominated accretion 
flow (CDAF). The dynamical 
structure of CDAFs is characterized by a 1/2 law of the radial 
density profile, $\rho \propto R^{-1/2}$,
shallower than that for the self-similar ADAF model,
$\rho \propto R^{-3/2}$,
where $\rho$ is the density and $R$ 
is the radius. In the limit of perfect self-similarity, CDAFs are 
nonaccreting with the radial 
velocity $\v = 0$ and the mass accretion rate $\dot M = 0$
(Narayan et al. 2000). 
In realistic CDAFs $\dot M$ is small but not exactly zero, leading to a
finite $\v \propto R^{-3/2}$.
The low luminosities of RIAFs are referred to a small convective luminosity,
$L_c = \varepsilon \dot M c^2$, with the convective efficiency
$\varepsilon \approx 0.01$ (Igumenshchev \& Abramowicz 2001), rather than
to the inward bulk advection of energies as in the ADAF model.

The self-similar CDAF model (as well as all other self-similar models),
though very clear and instructive, has its limitations. It is only
a local, not a global solution of a RIAF, in the 
sense that it can only be valid for the region of a RIAF far away
from the boundaries. In 
particular, it cannot reflect the transonic radial motion --
the most fundamental feature of black 
hole accretion flows. Advection ought to be important in the vicinity
of the black hole 
because of the large radial velocity of the accretion flow, 
so the inner region of RIAFs is 
likely to be better described by the ADAF model. 
Abramowicz et al. (2002) did suggest such 
a two zone structure of RIAFs: an outer convection-dominated zone 
and an inner 
advection-dominated zone, separated at a transition 
radius $\sim 50 R_g$ ($R_g = 2GM/c^2$ is the 
gravitational radius, with $M$ being the black hole mass).

In this paper we solve numerically the set of one dimensional 
(1D) height-integrated 
dynamical equations and obtain global solutions of RIAFs around
nonrotating black holes. 
Such a global solution is more exact and complete than the
self-similar solution on one hand, 
and is simpler and more transparent than the 2D or 3D simulations
on the other hand. 
Remember that the ADAF model was also in a self-similar form when
it was proposed (Narayan \& Yi 1994, 1995), and was later checked and
improved by authors working on the 1D global solution
(e.g. Narayan, Kato \& Honma 1997; Chen, Abramowicz \& Lasota 1997; 
Lu, Gu \& Yuan 1999).

\section{Equations}

We consider a set of stationary height-integrated equations describing
a RIAF around a 
nonrotating black hole (e.g. Narayan et al. 2000;
Abramowicz et al. 2002). In the absence of 
mass outflows, the continuity equation reads
\beq
\dot M = -2\pi R \Sigma \v = {\rm const},
\eeq
where $\Sigma = 2H\rho$ is the surface density, $H = c_s/\Omega_K$ is
the scale height, $c_s = (P/\rho)^{1/2}$ is the sound 
speed, $P$ is the pressure, $\Omega_K = (GM/R)^{1/2}/(R-R_g)$ is the
Keplerian angular velocity in the well known
Paczy\'{n}ski \& Wiita (1980) potential.

The radial momentum equation is as usual
\beq
\v \frac{d\v}{dR} + (\Omega_K^2-\Omega^2)R
+ \frac{1}{\rho}\frac{dP}{dR} = 0,
\eeq
where $\Omega$ is the angular velocity. Note that the ram-pressure
term $\v d\v /dR$ in equation (2) was ignored in the self-similar
CDAF model (Narayan et al. 2000), while we include it here in 
order to have a global solution.

In the presence of convection, the angular momentum and energy
equations can be written as
\beq
J = J_{adv} + J_{vis} + J_{con} = {\rm const},
\eeq
and
\beq
F = F_{adv} + F_{dis} + F_{con} = {\rm const},
\eeq
where $J$, $J_{adv}$, $J_{vis}$ and $J_{con}$ are the total, advective,
viscous and convective angular momentum 
flux, and $F$, $F_{adv}$, $F_{dis}$ and $F_{con}$ are the total,
advective, dissipative and convective energy flux, 
respectively. In the angular momentum equation (3),
advection moves angular momentum 
inward ($\v < 0$),
\beq
J_{adv} = 2\pi R \Sigma \v (\Omega R^2).
\nonumber
\eeq
Normal viscosity transports angular momentum outward, i.e. 
the viscous angular momentum 
flux is oriented down the angular velocity gradient, 
\beq
J_{vis} = -2\pi R \nu \Sigma R^2 (d\Omega / dR),
\nonumber
\eeq
where $\nu$ is the kinematic viscosity coefficient,
$\nu = \a c_s^2/\Omega_K$, with $\a$ being the constant 
Shakura-Sunyaev parameter. The basic question of how convection 
transports angular 
momentum is a complex topic. As discussed by Igumenshchev (2002), in
magnetohydrodynamical (MHD) CDAFs convection can transport angular
momentum either inward 
or outward, depending on the properties of turbulence in 
rotating magnetized plasma, which 
are not fully understood yet; but in HD
CDAFs we consider here, convection 
transports angular momentum inward, i.e. the convective flux 
is directed down the specific 
angular momentum gradient,
\beq
J_{con} = -2\pi R \nu_{con} \Sigma [d(\Omega R^2)/dR],
\nonumber
\eeq
where $\nu_{con}$ is the diffusion coefficient. In the energy
equation (4) the advective energy flux is 
\beq
F_{adv} = 2\pi R \Sigma \v B,
\nonumber
\eeq
where
$B = 0.5\v^2 - GM/(R-R_g) + 0.5R^2\Omega^2 + \gamma c_s^2/(\gamma-1)$
is the Bernoulli function, with $\gamma$ being the adiabatic index.
The dissipative energy flux is due to both the viscous and the 
convective shear stress, 
\beq
F_{dis} = \Omega (J_{vis} + J_{con}).
\nonumber
\eeq
The convective energy flux can be expressed in the form,
\beq
F_{con} = -2\pi R \nu_{con} \Sigma T (ds/dR),
\nonumber
\eeq
where $s$ is the specific entropy and $T$ is the temperature, and
$T\frac{ds}{dR} \equiv \frac{1}{\gamma-1}\frac{dc_s^2}{dR} -
\frac{c_s^2}{\rho}\frac{d\rho}{dR}$.

The same $\nu_{con}$ appears in both the expression for $J_{con}$
and that for $F_{con}$. This means that we have adopted the assumption
of Narayan et al. (2000) that all transport phenomena due to 
convection have the same diffusion coefficient, which is defined as
\beq
\nu_{con} = (L_M^2/4)(-N_{eff}^2)^{1/2},
\eeq
where $L_M = 2^{-1/4} l_M H_P$ is the characteristic mixing length,
$l_M$ is the dimensionless mixing-length parameter
(taken to be equal to $\sqrt{2}$ in our calculations),
$H_P = -dR/d\ln P$ is the pressure scale height, and
$N_{eff}$ is the effective frequency of convective blobs,
\beq
N_{eff}^2 = N^2 + \kappa^2,
\eeq
with $N$ and $\kappa$ being the Brunt-V\"{a}is\"{a}l\"{a} frequency
and the epicyclic frequency, respectively,
\beq
N^2 = -\frac{1}{\rho}\frac{dP}{dR}\frac{d}{dR}\ln
\left( \frac{P^{1/\gamma}}{\rho} \right) ,
\nonumber
\eeq
and
\beq
\kappa^2 = 2\Omega^2 \frac{d\ln (\Omega R^2)}{d\ln R} .
\nonumber
\eeq
Note that $\kappa \neq \Omega$ in general, $\kappa = \Omega$ only
for the self-similar scaling $\Omega \propto R^{-3/2}$
(Narayan et al. 2000). Convection is present whenever $N_{eff}^2 < 0$.
$\nu_{con}$ can also be written in the form similar to normal viscosity,
\beq
\nu_{con} = \ac c_s^2/\Omega_K ,
\eeq
where $\ac$ is a dimensionless parameter that describes the strength
of convective diffusion, it 
is not a constant, whereas the Shakura-Sunyaev viscosity
parameter $\a$ is assumed to be.

The angular momentum equation (3) can be rewritten explicitly as
\beq
\frac{(\a + \ac) c_s^2 R^2}{\Omega_K \v} \frac{d\Omega}{dR}
= \Omega R^2 \left( 1-\frac{2\ac c_s^2}{R\Omega_K \v} \right) - j ,
\eeq
where $j$ is a constant. We impose a no-torque condition
$d\Omega /dR = 0$
at the inner boundary of the accretion flow, so $j$ represents the
specific angular momentum accreted by the black hole in 
the absence of convection (i.e. when $\ac = 0$). The differential
form of the energy equation (4) is
\beq
\Sigma \v T \frac{ds}{dR} = \frac{c_s^2}{\Omega_K}\Sigma R
\left[ (\a +\ac) R \left( \frac{d\Omega}{dR} \right)^2 +
2\ac \Omega \frac{d\Omega}{dR} \right] 
+ \frac{1}{R} \frac{d}{dR} \left( R\frac{\ac c_s^2}{\Omega_K}\Sigma T
\frac{ds}{dR} \right) .
\eeq

Equations (1), (2), (8) and (9) can be solved for four
variables $\rho$, $\v$, $c_s$ and $\Omega$ as functions of 
$R$, provided the constant flow parameters $M$, $\dot M$, $\a$, $\gamma$ 
and $j$ are given. Note that $\ac$ is not another unknown quantity,
it can be obtained self-consistently from equations (5) and (7)
if $N_{eff}^2$ calculated from equation (6) is negative
(i.e. there is convection), otherwise it is zero (i.e. no convection).

\section{Global solutions}

We use the standard Runge-Kutta method to solve the set of three
differential equations (2), (8) and (9) for three unknowns
$\v$, $c_s$ and $\Omega$, and then obtain $\rho$ from 
equation (1). We integrate the 
differential equations from the sonic point $R_s$ (where $|v| = c_s$) 
both inward and outward. As discussed in detail by Abramowicz et al.
(2002), $R_s$ is not an additional free parameter, it is an 
eigenvalue and is self-consistently determined in a regular 
transonic solution. At and inside $R_s$ 
the flow ought to be advection-dominated, and convection is 
unimportant. So we set $\ac = 0$ in equations (8) and (9) when
starting the integration from $R_s$. The inward, supersonic part of 
the solution extends to the inner boundary of the flow, i.e. 
to a radius where the no-torque condition $d\Omega /dR = 0$
(i.e. $\Omega R^2 = j$) is satisfied. More important 
for our purpose here is the outward, subsonic part of the solution.
The Runge-Kutta method does not require any a priori 
outer boundary conditions, we just observe how the outward solution 
evolves with increasing $R$ until a reasonable outer boundary is found.
Whether and where there is convection in the 
flow are judged in the following self-consistent manner: at each 
radius we calculate $N_{eff}^2$ from equation (6), if
$N_{eff}^2 \ge 0$, i.e. no convection develops, then $\ac$
keeps to be zero; when $N_{eff}^2 < 0$, i.e. convection is present,
we obtain a non-zero $\ac$ from equations (5) and (7), and put it 
into equations (8) and (9) for the next step of the outward integration.

\begin{figure}
\centering
\includegraphics[width=16cm]{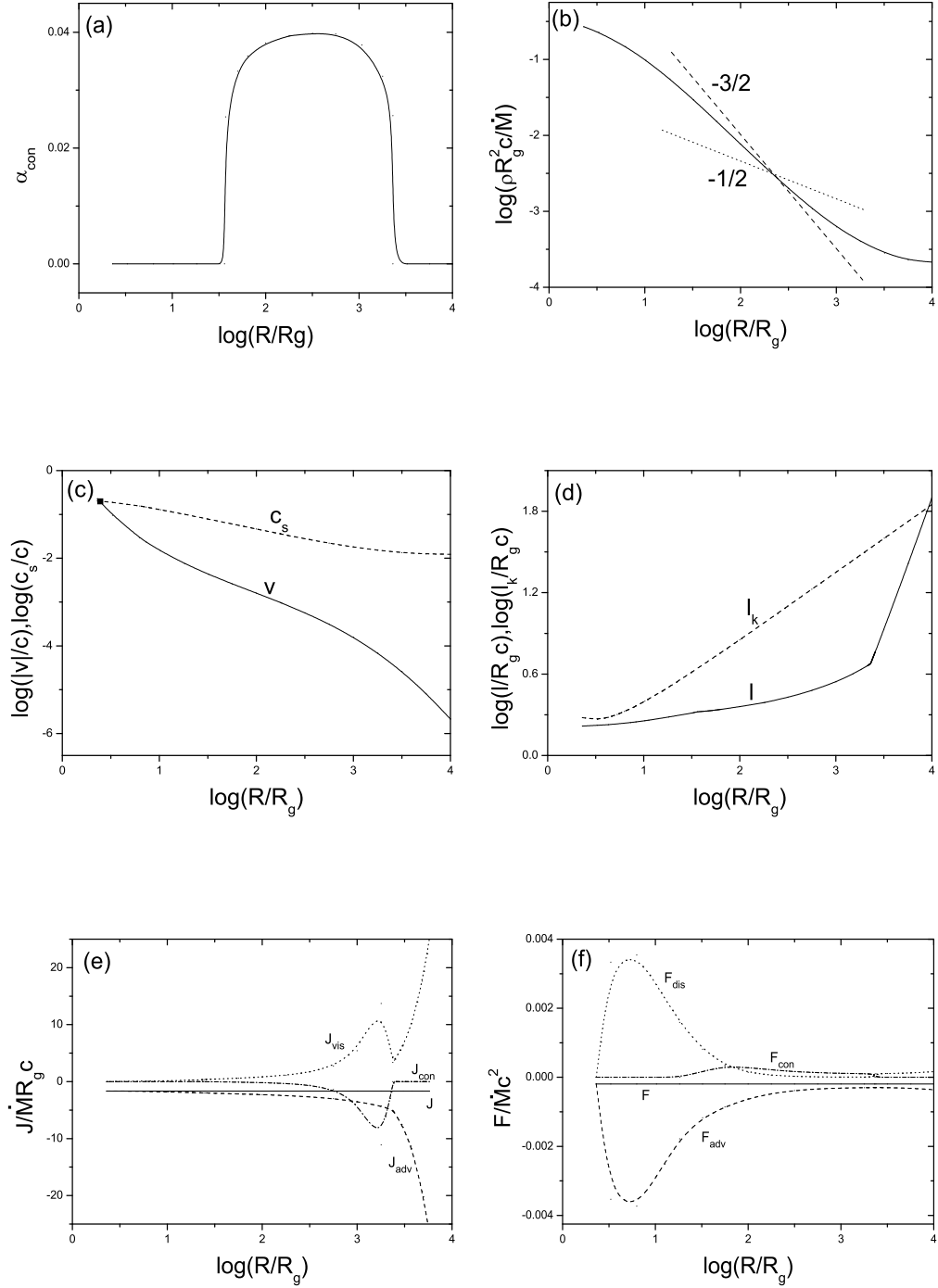}
\caption{
Example of type II three-zone solution for the viscosity
parameter $\a = 0.01$. (a), (b), (c), (d), (e) and (f) are for the
convective diffusion parameter $\ac$; the density $\rho$; the radial
velocity $\v$ and the sound speed $c_s$; the specific angular momentum
$l$ and the Keplerian angular momentum $l_K$; the total angular momentum
flux $J$ and its advective component $J_{adv}$, viscous component
$J_{vis}$ and convective component $J_{con}$; and the total energy flux
$F$ and its advective component $F_{adv}$, dissipative component
$F_{dis}$ and convective component $F_{con}$; respectively.
}
\end{figure}

We obtain three types of global solutions depending on the value of 
the viscosity parameter $\a$: \\
I. \ \ Pure ADAF solution for large $\a \ga 0.1$. In this case viscous
action is so strong that the flow is totally advection-dominated,
and no convection develops at all. This type of solution 
has been extensively investigated in the literature 
(e.g. Narayan et al. 1997), and we do not repeat it here. \\
II. \ \ Three zone solution for moderate $\a \sim 0.01$. Fig.~1 provides 
an example of this type of 
solution, with $\a = 0.01$, $\gamma = 5/3$, $j = 1.646(cR_g)$, and
$R_s = 2.3R_g$. Fig.~1(a) shows how the 
convective diffusion parameter $\ac$ varies with the radius $R$, 
from which a three-zone structure is clearly seen. In the middle zone
ranging from $R = 34R_g$ to $R = 2300R_g$, convection 
develops and plays a dominant role, in the sense that $\ac > \a$ 
for almost all the zone. For the 
inner zone ($R < 34R_g$) and the outer zone ($R > 2300R_g$) convection 
ceases to exist ($\ac = 0$). 
Fig.~1(b) draws the radial profile of the density $\rho$ (the solid line). 
For comparison, the profiles 
$\rho \propto R^{-3/2}$ of the self-similar ADAF solution
(the dashed line) and $\rho \propto R^{-1/2}$ of the 
self-similar CDAF solution (the dotted line) are also given. It is 
seen that the radial density 
distribution in the global solution is in between, i.e. shallower 
than that for the self-similar 
ADAF solution, and steeper than that for the self-similar CDAF 
solution. Fig.~1(c) is for the 
radial velocity $\v$ (the solid line) and the sound speed $c_s$ 
(the dashed line). The sonic point is 
marked by a filled square. Fig.~1(d) is for the specific 
angular momentum $l = \Omega R^2$ (the solid 
line) and the Keplerian angular momentum $l_K = \Omega_K R^2$
(the dashed line). The profiles of $l$ in the 
three zones are distinct from each other: in the inner zone $l$ 
behaves as in the pure ADAF 
solution of type I; in the middle zone the profile is greatly 
flattened comparing with the pure 
ADAF solution would have, because of the strong inward transport of 
angular momentum by 
the convective flux; while in the outer zone $l$ increases steeply 
with increasing $R$, and reaches 
the Keplerian value $l_K$ at $R = 9252R_g$, and this radius can be 
reasonably regarded as the outer 
boundary of the flow. Fig.~1(e) is devoted to the angular momentum 
flux, in which the total 
flux $J$, the advective component $J_{adv}$, the viscous component
$J_{vis}$ and the convective component $J_{con}$ are denoted by the solid,
dashed, dotted and dot-dashed line, respectively. In 
the inner advection-dominated zone $J_{con} = 0$, and $J_{vis}$ is
very small, so $J$ is dominated by $J_{adv}$. 
In the middle convection-dominated zone $J_{con}$ (inward) and 
$J_{vis}$ (outward) almost cancel each 
other, while $J_{adv}$ is relatively small. In the outer no-convection 
zone $J_{con}$ becomes zero again, 
and both $J_{vis}$ and $J_{adv}$ (its absolute value) increase with 
increasing $R$. The competition of these 
three components results in a constant net flux throughout the 
flow, $J = - \dot M j = -1.646(\dot M cR_g)$, which is inward.
In Fig.~1(f) which is for the energy flux, the total flux $F$, the 
advective component $F_{adv}$, the dissipative component $F_{dis}$
and the convective component $F_{con}$ 
are denoted again by the solid, dashed, dotted and dot-dashed line,
respectively. In the inner and the outer
zone $F_{con} = 0$, and $F_{adv}$ (inward) and $F_{dis}$ (outward) almost 
balance in power. In the middle 
zone $F_{dis}$ is small, and $F_{adv}$ is nearly balanced by 
$F_{con}$ (outward) instead. The net result of the 
competition is again a constant flux $F = -0.00019(\dot M c^2)$, 
with the efficiency $\varepsilon = 0.00019$. 
Note that although $F_{con}$ alone is positive, $F$ is negative, i.e. 
the released gravitational energy is 
dragged inward as in ADAFs, and that the efficiency of energy 
release is very small. \\
\begin{figure}
\centering
\includegraphics[width=16cm]{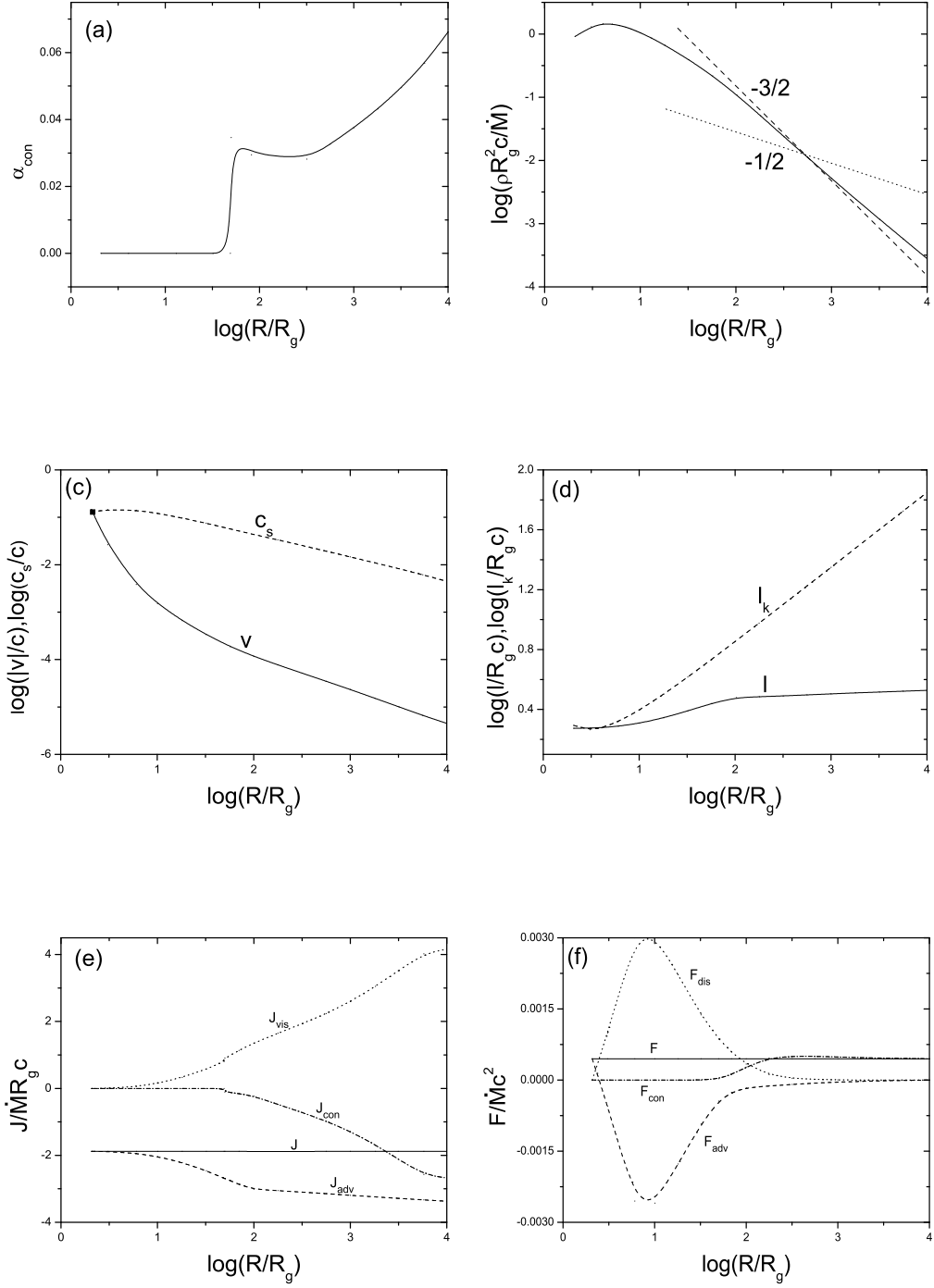}
\caption{
Example of type III two-zone solution for $\a = 0.001$.
The arrangements are the same as for Fig.~1.
}
\end{figure}
III. \ \ Two zone solution for small $\a \la 0.001$. This type of 
solution has been suggested 
previously by Abramowicz et al. (2002), of which an example 
is given by Fig.~2, with $\a = 0.001$,
$\gamma = 5/3$, $j = 1.88(cR_g)$, and $R_s = 2.1R_g$. The arrangements 
and the symbols of the figure 
are the same as for Fig.~1. The solution has a two-zone 
structure, i.e. an inner 
advection-dominated zone and an outer convection-dominated zone, 
with a transition 
occurring at $R = 49R_g$. It is seen from Fig.~2(a) that $\ac = 0$ 
in the inner zone, it is non-zero 
and increases with increasing $R$ in the outer zone, i.e. convection 
does not cease to exist for 
large radii as in type II solution. In Fig.~2(b) the radial 
density profile is also between the line 
of self-similar ADAF solution ($\propto R^{-3/2}$) and that of 
self-similar CDAF solution ($\propto R^{-1/2}$). In 
Fig.~2(c) the profile of $\v$ in the outer zone is shallower than 
that in Fig.~1(c), proving a 
stronger effect of convective motion against advection. The power 
of convection is most 
clearly seen from Fig.~2(d): in the outer zone the inward transport 
of angular momentum by 
convection is so effective that $l$ keeps being almost constant. 
Accordingly, the combined 
inward flux of $J_{con}$ and $J_{adv}$ overcomes the outward flux $J_{vis}$, 
resulting in an inward net flux $J = -1.88(\dot M cR_g)$, as drawn
in Fig.~2(e). Fig.~2(f) is noticeable, as it shows that in the outer zone
both $F_{adv}$ and $F_{dis}$ tend to be zero for large radii, and the
outward $F_{con}$ really dominates, giving a net energy flux which is
positive, $F = 0.00045(\dot M c^2)$, with $\varepsilon = 0.00045$.
As Abramowicz et al. (2002) argued but not explicitly proved,
the outward $F$ is produced in the inner zone where most of the
dissipatively released gravitational energy (i.e. $F_{dis}$)
is advected inward (i.e. $F_{adv}$), 
with a small remainder that bubbles out through the flow. 
This outward $F$ is a characteristic feature of this type of solution,
qualitatively different from the case of ADAFs. 

Figs.~1 and 2 are for $\gamma = 5/3$. We have also made calculations
for different values of $\gamma (4/3 \leq \gamma \leq 5/3)$, and the
results obtained remain qualitatively similar.

\section{Discussion}

We have shown that global solutions of black hole RIAFs can be divided
into three types according to the strength of normal viscosity.
When viscosity is strong (large $\a$), convection 
plays no role, and the flow is totally advection-dominated 
(type I solution). If viscosity is 
moderate (smaller $\a$), the flow has a three-zone structure, 
and convection is important only in 
the middle zone which ranges from a few tens to a few thousands 
of $R_g$; the net energy flux is 
still inward as in ADAFs (type II solution). In the case of weak 
viscosity (very small $\a$), the 
flow consists of two zones with a transition radius of a few 
tens of $R_g$, and convection 
dominates in the outer zone; the net energy flux becomes 
outward (type III solution). Our type 
III solution confirms the idea of two-zone structure proposed 
by Abramowicz et al. (2002), 
though the transition radius is defined in somewhat different ways.

Our results are in good agreement with those of numerical simulations,
and may be somewhat detailed improvements on the self-similar CDAF model.
Here we address a few points: \\
1. As Igumenshchev \& Abramowicz (2001) summarized, 2D HD
simulations of RIAFs had proven that for $\a \la 0.03$ and all the
reasonable values of $\gamma$ the flow is convectively unstable, which
agrees very well with our results here; and that convection 
transports angular momentum inward and no outflows are present, 
which support our assumptions of inward flux $J_{con}$ and constant
accretion rate $\dot M$. \\
2. In the self-similar CDAF model convection was assumed a priori to 
present throughout the flow, and the convective diffusion 
parameter $\ac$ was treated as a constant (Narayan et al. 2000);
while in our global solutions whether and where convection develops
in a given flow are self-consistently determined by 
calculating the effective frequency $N_{eff}$ at each radius, and 
$\ac$ is a function of $R$ and is also calculated. \\
3. In our solutions the density profile is between the 1/2 law of
self-similar CDAF solution and 3/2 law of self-similar ADAF solution. 
We think this is reasonable, because the self-similar CDAF and ADAF 
solutions should be regarded as two ideal extremes, and should 
be modified under the influence of boundary conditions in global solutions.
In their most recent 3D MHD simulations of RIAFs,
Pen, Matzner \& Wong (2003) found a quasi-hydrostatic density profile
$\rho \propto R^{-0.72}$ (i.e. also between the 1/2 and 3/2 laws). \\
4. As Pen et al. (2003) pointed out, the 1/2 law derives from assuming
a positive convective energy flux $F_{con}$. In our solutions although
$F_{con} > 0$ holds, the total energy flux $F$ can be either 
positive (type III solution) or negative (type II solution),
depending on whether convection really dominates. In fact
Pen et al. (2003) also obtained $F < 0$, which is consistent with our 
type II solution. \\
5. The efficiency of energy release of RIAFs was estimated previously
as $\varepsilon \approx 0.003-0.01$ (e.g. Igumenshchev \& Abramowicz 2000; 
Abramowicz et al. 2002), while in our solutions $\varepsilon$ 
is significantly smaller (it is 0.00019 in Fig.~1 and 0.00045 in Fig.~2).
This is probably because those authors referred $\varepsilon$ only to the 
convective energy flux $F_{con}$ (they named $\varepsilon$
'convective efficiency'), while we refer it to the 
total energy flux $F$. The extremely low efficiency in our solutions
might have observational implications, e.g. it might help explain 
the immense discrepancy between the dynamically estimated 
mass accretion rate and the observed luminosity in the Galactic center
and other nearby galaxies (e.g. Pen et al. 2003).

Concerning these main results, there are also several points we need to
comment on, only very briefly: \\
1. Our treatment of convection here is appropriate for viscous HD flows
and not necessarily for MHD ones. Although since the work of Balbus \&
Hawley (1991) it has become widely agreed that the magneto-rotational
instability is the detailed mechanism that produces viscosity, it is
not yet clear how the MHD approach could change the results of the HD model.
For example, depending on the assumed topology of the magnetic field in
the flow, some studies find quite good agreement between numerical MHD
simulations and analytical works on HD CDAFs (e.g. Machida, Matsumoto \&
Mineshige 2001); others, however, claim that there are significant
differences (e.g. Hawley \& Balbus 2002). It is therefore not
surprising that we compare our results here with not only HD, but also
some MHD simulations. \\
2. We identify 5 controlling parameters of the flow, namely
$M$, $\dot M$, $\a$, $\gamma$ and $j$, from which
the detailed flow structure is determined. Of these parameters,
$M$, $\dot M$ and $j$ scale variables $R$, $\rho$,
$l$, $J$ and $F$, then our numerical calculations show that $\a$
is the only important one in determining the solution topology,
and $\gamma$ seems to be insignificant. These results are
consistent with those of previous 2D HD simulations. For example,
Fig.~1 of Igumenshchev \& Abramowicz (2001) shows clearly that for
small $\a \la 0.03$, flows are convectively unstable
regardless of the value of $\gamma$; this agrees well with our type
II and III solutions. However, the same figure also indicates that in
the case of large $\a \ga 0.1$, $\gamma$ is important in determining
other properties of flows such as large-scale circulations and outflows;
this would complicate our type I solution, but has gone beyond
the scope of the present paper. \\
3. We use $N_{eff}^2 < 0$ for the onset of convection. One might wonder
if this is a sufficient criterion. Of course the best way of verification
is numerical simulations. Another simpler means is to estimate the
Rayleigh number defined as $R_a = g a \Delta T R^3 / \nu ^2$,
taking the thermal diffusivity to equal the kinematic viscosity $\nu$
(i.e. the Prandtl number $P_r = 1$), where $g$ is the
specific gravitational force, and $a$ is the volume expansion coefficient.
Substituting $g \sim GM/R^2$, $\nu = \a c_s H = \a c_s^2 / \Omega_K$, and
$\Omega_K^2 \sim GM/R^3$, one has $R_a \sim a \Delta T / (H/R)^4 \a ^2$.
Now $a \sim 10^{-3}-10^{-4}$ for gaseous materials, and $\Delta T$ must
be more than enough to make ($a \Delta T$) larger than
unity, then for moderate values $(H/R) \sim 0.1$ and $\a \sim 0.01$,
$R_a$ safely exceeds 1000, the critical value required by laboratory
convection. This argument further implies that convection is likely
to develop in the radial direction of the flow. \\
4. In our solutions the very low efficiency of energy release $\varepsilon$
corresponds to the net energy flux $F$ that results from the competition
of the advective, dissipative, and convective components. For a plain RIAF,
i.e. that with zero radiative cooling, $F$ flows constantly either inward
toward the central black hole ($F < 0$, type II solution) or outward
toward the outer boundary of the flow ($F > 0$, type III solution).
If a RIAF is not plain, i.e. a very small but non-zero amount of energy is
radiated away, which is more likely to be the realistic case, then $F$
provides energy available for radiation, and $\varepsilon$ gives an
estimation of the radiative efficiency. Certainly, this very small
radiative loss of energy cannot affect the dynamics of the flow.

\section*{Acknowledgments}
This work is supported by the National Science Foundation of China
under Grant No.10233030.

\end{document}